\documentclass[a4paper,fleqn,usenatbib]{mnras}
\usepackage{savesym}
\usepackage[T1]{fontenc}
\usepackage{ae,aecompl}

\usepackage{graphicx}
\usepackage{amsmath}
\usepackage{amssymb}
\savesymbol{iint}
\usepackage{txfonts}
\restoresymbol{TXF}{iint}

\title[Neural networks]{Asteroseismic determination of fundamental parameters of sun-like stars using multi-layered neural networks}

\author[Verma et al.]
{\parbox{\textwidth}{Kuldeep Verma,$^{1}$\thanks{E-mail: kuldeepv@tifr.res.in}
Shravan Hanasoge,$^{1}$
Jishnu Bhattacharya,$^{1}$
H. M. Antia$^{1}$
and \,\,\,\,\,Ganapathy Krishnamurthi$^{2}$}\vspace{0.4cm}
\\
$^{1}$Tata Institute of Fundamental Research, Homi Bhabha Road, Mumbai 400005, India\\
$^{2}$Department of Engineering Design, IIT Madras, Chennai-36, India
}

\date{Accepted XXX. Received YYY; in original form ZZZ}
\pubyear{2016}

\begin{document}
\label{firstpage}
\pagerange{\pageref{firstpage}--\pageref{lastpage}}
\maketitle

\begin{abstract}
The advent of space-based observatories such as CoRoT and {\it Kepler} has enabled the testing of our understanding of stellar evolution on thousands of stars. Evolutionary models typically require five input parameters, the mass, initial Helium abundance, initial metallicity, mixing-length (assumed to be constant over time), and the age to which the star must be evolved. Some of these parameters are also very useful in characterizing the associated planets and in studying galactic archaeology. How to obtain these parameters from observations rapidly and accurately, specifically in the context of surveys of thousands of stars, is an outstanding question, one that has eluded straightforward resolution. For a given star, we typically measure the effective temperature and surface metallicity spectroscopically and low-degree oscillation frequencies through space observatories. Here we demonstrate that statistical learning, using artificial neural networks, is successful in determining the evolutionary parameters based on spectroscopic and seismic measurements. Our trained networks show robustness over a broad range of parameter space, and critically, are entirely computationally inexpensive and fully automated. We analyze the observations of a few stars using this method and the results compare well to inferences obtained using other techniques. This method is both computationally cheap and inferentially accurate, paving the way for analyzing the vast quantities of stellar observations from past, current, and future missions.
\end{abstract}

\begin{keywords}
stars: fundamental parameters---stars: interiors---stars: low-mass---stars: oscillations---stars: solar-type
\end{keywords}

\section{Introduction}
\label{sec:intro}
Tracing the narratives of stellar lives is a problem of fundamental importance. The vast repositories of stellar observations allow for precision testing of our understanding of stellar evolution. CoRoT \citep{bagl09} and {\it Kepler} \citep{boru09} have together taken observations of oscillations of thousands of stars across the Hertzsprung-Russell diagram. Upcoming space observatories, such as TESS \citep{rick14} and PLATO 2.0 \citep{raue14}, will take measurements of light curves of possibly hundreds of thousands of stars. However, analyzing such large observational datasets and characterizing the stars is challenging.

Observations of stellar oscillations provide strong constraints on the properties of stellar interior, enabling the determination of fundamental stellar parameters, e.g. the mass, radius, and the age, to unprecedented precision. For instance, \citet{math12} performed an asteroseismic study on 22 of the brightest solar-type stars observed by {\it Kepler}, and inferred the masses and the radii at a precision level of 1\% and ages to within 2.5\% \citep[see also,][]{metc14,metc15}. The extraordinary precision obtained by \citet{math12} might be deceptive as they did not incorporate the uncertainties in stellar physics in their error estimation. In a thorough study of the impact of uncertainties in stellar physics, \citet{lebr14} performed a case study on a planet-host star, HD52265, observed by CoRoT satellite. They optimized the stellar parameters for various combinations of the oscillation frequencies, large and small frequency separations, frequency ratios, and spectroscopic observables with different input physics, and found the mass, radius, and age of HD52265 with uncertainties of 7\%, 1.5\%, and 10\%, respectively. \citet{silv15} applied a Bayesian scheme to grids of evolutionary models and found the masses, radii, and ages of 33 {\it Kepler} planet-host stars with median statistical uncertainties of 3.3\%, 1.2\%, and 14\%, respectively \citep[see also,][]{chap14}. Recently, \citet{rees16} carried out a hare-and-hounds exercise to evaluate the accuracy of seismic determinations and the reliability of associated error bars \citep[see also,][]{vall14}. They used different approaches and evolutionary codes to derive the stellar parameters using the simulated seismic and spectroscopic data, and found that, on average, the masses, radii, and ages can be derived to a precision of 3.9\%, 1.5\%, and 23\%, respectively. The various algorithms satisfy the PLATO specifications for precision levels on the inference of mass and radius and is close for the age.

It is important to accurately determine the fundamental parameters of host stars in order to accurately characterize the associated planets. There are several individual planet-host stars studied using asteroseismology \citep[see, e.g.,][etc.]{gill11,nutz11,esco12,gill13,lebr14}. \citet{hube13} obtained fundamental properties of 66 {\it Kepler} planet-candidate host stars using asteroseismology, which led to a better characterization of the associated planets. \citet{liu14} were able to better constrain the properties of exoplanets based on inferences of parameters of the corresponding host stars for six stellar systems. \citet{silv15} presented a uniform study of 33 {\it Kepler} planet-candidate host stars and derived the stellar properties with high precision. The fundamental stellar parameters are also used to study stellar populations in the Milky Way. For example, \citet{chap11} performed an ensemble study on 500 solar-type stars observed by {\it Kepler} and found that the distribution of observed masses differ from the predictions of the models of synthetic stellar populations in the Milky Way. The stellar properties are also used to study the history of our galaxy \citep{migl13,casa14}. 

The connection between the stellar parameters (the mass $M$, initial Helium abundance $Y_i$, initial metallicity $Z_i$, mixing-length $\alpha_{\rm MLT}$, and the age $t$) and the observables (the effective temperature $T_{\rm eff}$, surface metallicity $[{\rm Fe}/{\rm H}]_s$, luminosity $L$, and the oscillation frequencies $\nu_{nl}$, where the indices $n$ and $l$ are radial order and spherical harmonic degree, respectively) is complicated. Although all these quantities are intimately connected through the physics of the problem, the problem of inference is hampered by the facts that there is no known analytical, direct inverse relation between them and the constellation of stellar parameters is likely connected to the observables in a highly nonlinear manner. Because of the former, attempts are made to sample the model space (using for instance Monte-Carlo simulations) in order to identify the most relevant swath of models. Further, because the inference consists of using tens of inputs and generating several outputs, the problem lives in a high-dimensional space. Retrieving the underlying non-linear relationships is a difficult problem, especially by conventional methods such as regression. Among the many approaches in use \citep[see, e.g.,][etc]{metc09,grub12,silv15,appo15}, a majority involve computing several evolutionary tracks for individual stars and optimizing the stellar parameters for a given set of observables. In addition to being cumbersome and computationally expensive, the associated uncertainties on the stellar parameters are difficult to estimate, particularly for methods that rely on local optimization algorithms \citep{rees16}. Simpler methods that rely on searching fixed grids are computationally efficient, but the accuracy of such techniques is limited by the resolution of the grid and the systematics associated with focusing on narrow search regions (most grids, e.g., the GARSTEC grid used in \citet{silv15}, use solar calibrated mixing-length and a Helium-to-metal enrichment law to estimate the $Y_i$). 

In this paper, we apply artificial neural networks to determine the stellar properties based on classical spectroscopic and seismic measurements accurately and rapidly.

\begin{figure}
\includegraphics[width=\linewidth]{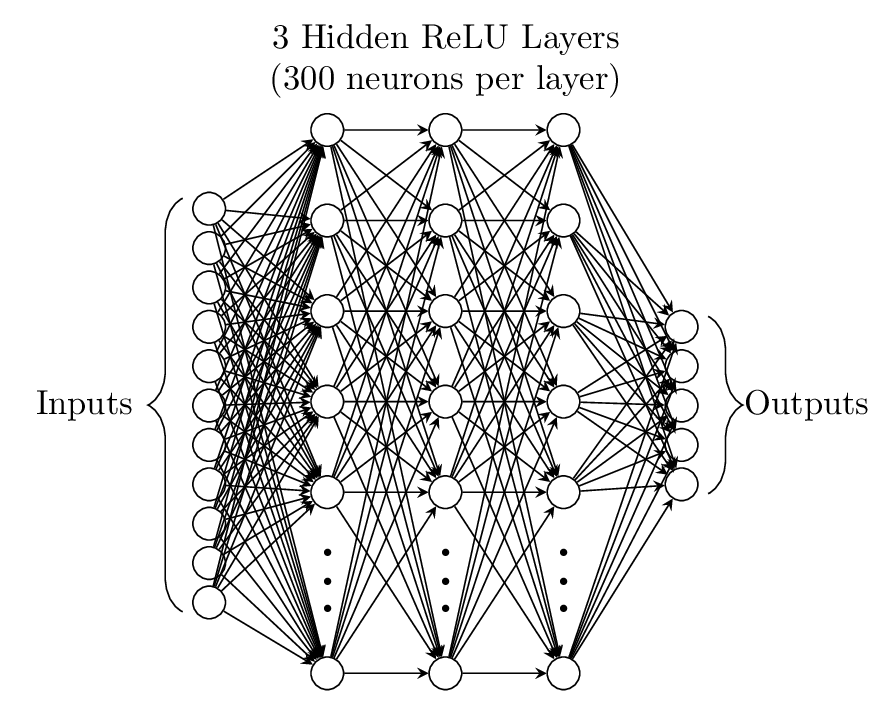}
\caption{Sketch of neural network with inputs and outputs. The output of a rectified linear unit (ReLU) is equal to the input if positive and otherwise zero.\label{fig1}}
\end{figure}

\section{Technique}
\label{sec:tech}
Neural Networks (NNs) have proven very successful at deciphering and isolating complicated and subtle correlations in large datasets. The prototypical NN consists of a layer of `neurons', each with inputs and outputs. These neurons communicate with each other, their actions defined by their weights and biases that are gradually tuned through the process of statistical training. By degrees, the network maps out the probability density function of the high-dimensional space of stellar models. For a sample of inputs (spectroscopic + seismic), the well-trained network is able to identify the relevant region of the evolutionary parameter space based on its understanding of the probability density function. 

The structure of a neural network, as shown in Figure~\ref{fig1}, contains several layers of neurons \citep{lecu15,schm15}. The choices of how many neurons per layer and the number of layers are part of a set of parameters that are termed ``hyper-parameters". These parameters depend on the complexity of the problem and heuristics are used to determine them \citep{berg12}. The network in Figure~\ref{fig1} for instance has been obtained by optimizing the hyper-parameters on a fixed grid. 

The neural network must be trained before it can be used to solve a physical problem. We implement a learning algorithm that uses stellar models to train the network in such a way that it can predict the fundamental stellar parameters for a given set of measurements.

\subsection{Learning algorithm}
\label{subsec:learn}
The neurons in the network have intrinsic biases and the connections between neurons of consecutive layers have associated weights. We denote the bias of the $j^{\rm th}$ neuron in the $l^{\rm th}$ layer by $b^l_j$ and the weight associated between the $k^{\rm th}$ neuron in the $(l-1)^{\rm th}$ layer and $j^{\rm th}$ neuron in the $l^{\rm th}$ layer by $W^l_{jk}$. The output---also known as the activation---of the $l^{\rm th}$ layer can be computed in terms of the output of the $(l-1)^{\rm th}$ layer using,   
\begin{equation}
a^l = \sigma(W^l a^{l-1} + b^l),
\end{equation}
where $\sigma$ is the activation function which depends on the type of neuron (sigmoid neuron, hyperbolic tangent neuron, rectifying neuron, etc.). In this paper, we have chosen the rectifying neurons, as they have been recently shown to be a better model of the biological neurons and yield equal or better performance than the sigmoid and hyperbolic tangent neurons \citep{glor11}. The output of the network $a^L = y$ can be computed recursively in terms of its weights and biases---which are to be determined---for a given input vector $a^1 = x$. 

The weights and biases of the network are determined by a process called training. The training needs a dataset---which should ideally represent the whole space of the problem at hand---containing both the input vectors $x_j$ and the corresponding output vectors $y_j$. An input vector together with the corresponding output vector is called an example. We define a regularized cost function,
\begin{equation}
C = \frac{1}{mn} \sum_{jx_j} |y_j - a^L_j|^2 + \lambda \sum_{jkl} |W^l_{jk}|^2,
\end{equation}
where $m$, $n$, and $\lambda$ are the length of the output vector, number of training examples, and the regularization parameter, respectively. This function is minimized with respect to the weights and the biases of the network. Typically, we need a large training dataset for a practical problem, as a consequence, the conventional gradient descent methods to update the weights and biases are very expensive. Hence, stochastic gradient descent---which approximates the true gradient of the cost function by the gradient obtained using a randomly selected subset, known as a mini-batch, of the full dataset---is used \citep{bott10}. The gradient of the cost function with respect to the weights and biases can be obtained using back-propagation algorithm \citep{rume86,lecu98}, which is described below.
\begin{enumerate}
\item Set the input vector as activation $a^1$. 
\item For each $l$ = 2, 3, ..., $L$; compute the vectors $z^l = W^l a^{l-1} + b^l$ and activations $a^l = \sigma(z^l)$.
\item Compute the vector $\delta^L = \nabla_a{C} \odot \sigma'(z^L)$, where $\nabla_a{C}$ is the vector $\partial{C}/\partial{a^L_j}$, $\odot$ denotes Hadamard product and $\sigma'(z^L) = \partial{a^L_j}/\partial{z^L_j}$.
\item For each $l$ = $L - 1$, $L - 2$, ..., 2; compute vectors $\delta^l = ((W^{l+1})^T \delta^{l+1}) \odot \sigma'(z^l)$.
\item Compute ${\partial C}/{\partial w^l_{jk}} = a^{l-1}_k \delta^l_j$ and ${\partial C}/{\partial b^l_j} = \delta^l_j$.
\end{enumerate}

\begin{figure}
\centering
\includegraphics[scale=0.5]{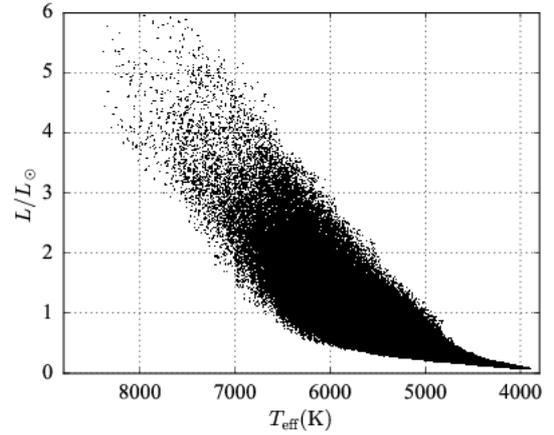}
\caption{Hertzsprung-Russell diagram showing the parameter space of the models used in training. Note that we focus here on solar-like main-sequence stars; in principle, the network can be trained on a variety of other types of stars as well.\label{fig2}}
\end{figure}

\subsection{Training models}
\label{subsec:train}
We compute the training models using the Modules for Experiments on Stellar Astrophysics code \citep[MESA;][]{paxt11,paxt13}. We use the OPAL equation of state \citep{roge02}, Opacity Project (OP) high-temperature opacities \citep{badn05,seat05} supplemented with low-temperature opacities from \citet{ferg05}. Metallicity mixtures from \citet{gs98} are used. We use nuclear reaction rates from NACRE \citep{angu99} for all reactions except $^{14}{\rm N}(p,\gamma)^{15}{\rm O}$ and $^{12}{\rm C}(\alpha,\gamma)^{16}{\rm O}$, for which updated reaction rates from \citet{imbr05} and \citet{kunz02} are used. Convection is modeled using the standard mixing-length theory \citep{cox68} without overshoot, and diffusion of Helium and heavy elements is incorporated using the prescription of \citet{thou94}.   

We construct 20,000 evolutionary tracks with initial parameters distributed randomly and uniformly in the following ranges: $M \in 0.70 - 1.10 M_\odot, Y_i \in 0.20 - 0.40, Z_i \in 0.003 - 0.040,$ and $\alpha_{\rm MLT} \in 1.20 - 2.50$. An equivalent Cartesian grid would on average have 12 equally spaced points ($12^4 = 20,736 \approx 20,000$) for each parameter, which corresponds to the spacings of approximately $0.033 M_\odot$, $0.0167$, $0.003$, and $0.11$ for $M$, $Y_i$, $Z_i$, and $\alpha_{\rm MLT}$, respectively. The randomly and uniformly distributed grid points are expected to sample the parameter space better. The stellar models are evolved and retrieved at regular intervals to cover a range of ages (on average, 18 models per evolutionary track). The evolution is stopped when the central Hydrogen mass fraction falls below $10^{-4}$ or the age exceeds 20 Gyr. We generate 360,000 models, spanning a range of masses, initial compositions, mixing lengths, and ages, and are shown in Figure~\ref{fig2}. We do not consider the mass range $1.10 - 1.60 M_\odot$ here (will be studied in the future) for the reasons that these stars have convective cores and a sixth parameter, the overshoot, has to be included. Furthermore, the current models of atomic diffusion are more subtle in this mass range and a careful treatment is required.

The adiabatic oscillation frequencies for all the models are obtained using the Adiabatic Pulsation code \citep[ADIPLS;][]{jcd08}. Stellar oscillation frequencies constrain fundamental stellar parameters \citep[see, e.g.,][]{brow94,jcd04,aert10}. The global properties of the power spectrum, namely the large frequency separation, $\Delta\nu_{n,l} = \nu_{n,l} - \nu_{n-1,l}$, and the frequency at which the power attains its maximum, $\nu_{\rm max}$, may be used to directly infer the mass and radius of the star using established scaling relations \citep{kjel95} to a precision of 10\% and 5\%, respectively \citep[see, e.g.,][]{hube11,hube12,silv12,epst14}. The small frequency separation, $\delta\nu_{n,l} = \nu_{n,l} - \nu_{n-1,l+2}$, depends on the sound-speed gradient in the core, and therefore, traces the evolutionary stage of the star \citep[see, e.g.,][]{ulri88,jcd88}. It has been shown that plotting the small and large frequency separations against each other (the ``asteroseismic diagram'') enables the determination of the masses and ages of stars with known compositions \citep[see Figure~1 of][]{jcd93}. Figure~1 of \citet{lebr14} shows a relatively modern version of the asteroseismic diagram. 

The theoretical model frequencies have systematic deviation from the observed frequencies due to the poor modeling of the near surface region \citep[see, e.g.,][]{jcd88b,dzie88,jcd96,jcd97}. This offset, known as the ``surface effect'', renders direct comparisons of the model and observed frequencies ambiguous. In the current problem, if we use the raw model frequencies to train the network,
the surface term will introduce systematic errors leading to incorrect prediction of stellar parameters for real stars. Therefore, we do not use the individual oscillation frequencies to train the network, and instead use the so-called ``frequencies ratios'' which are shown to be independent of the near surface layers \citep{roxb03,roxb05,oti05}. The two point and five point ratios are defined as,
\begin{equation}
r_{02}(n) = \frac{\nu_{n,0} - \nu_{n-1,2}}{\nu_{n,1} - \nu_{n-1,1}}
\end{equation}
and
\begin{equation}
r_{01}(n) = \frac{\nu_{n-1,0} - 4 \nu_{n-1,1} + 6 \nu_{n,0} - 4 \nu_{n,1} + \nu_{n+1,0}}{8(\nu_{n,1} - \nu_{n-1,1})}.
\end{equation}
We identify eleven observables from above models to be used as the input of the network ($[{\rm Fe}/{\rm H}]_s$, $T_{\rm eff}$, $L$, $r_{02}$ at radial orders 16--19, $r_{01}$ at orders 17--19, and the average large frequency separation $\langle\Delta\nu\rangle$) and five stellar parameters to be used as the output ($M$, $Y_i$, $Z_i$, $\alpha_{\rm MLT}$, and $t$). The network can be easily retrained for different numbers and ranges of radial orders ($\sim 50$ cpu hours for retraining). 

The whole set of 360,000 models are divided into three groups: training examples - 240,000 models, validation examples - 60,000 models, and test examples - 60,000 models. Training examples are used to define and tune the network parameters, i.e., the weights and biases, validation examples are used to tune the network hyper-parameters, i.e., the number of hidden layers, number of neurons per layer, mini-batch size, learning rate, regularization parameter etc., and the test examples are used to determine the accuracy of the predictive algorithm. We normalize the inputs and the outputs of the network by subtracting the corresponding mean values---obtained from the training data---and dividing the residual by the standard deviations. The original inputs and outputs can be recovered---when required---using the normalized values, mean values, and the standard deviations. In principle, the normalization is not necessary for this problem, however in practice, it makes training faster and reduces the chances of getting stuck in local optima. 

\begin{figure}
\includegraphics[width=\linewidth]{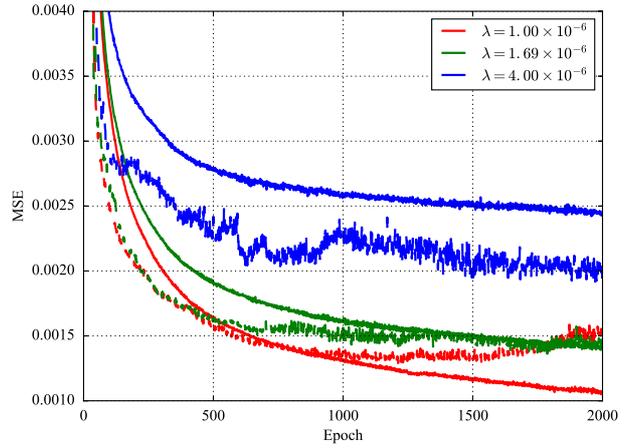}
\caption{Mean square errors as a function of epoch for three different choices of the regularization parameter; continuous and dashed lines correspond to the MSEs for the training and the validation data, respectively.\label{fig3}}
\end{figure}

\section{Results}
\label{sec:res}
The network training was performed using Theano \citep{berg10,bast12}, publicly available Python-based machine-learning software, which was run on the SEISMO cluster at TIFR. We used a mini-batch size of 50 for the computation of the gradient of the cost function to update the weights and biases using the stochastic gradient decent method. A too small mini-batch size results in a large error in the gradient computation, while a large mini-batch slows down the learning, and an optimal mini-batch size has to be found and used. The weights and biases were updated for every mini-batches of the training data (240,000/50 = 4,800 mini-batches $\equiv$ an `epoch'), and the process is repeated for several epochs until the convergence is achieved. The validation data are used to check the sanity of the training. For example, Figure~\ref{fig3} shows the mean square errors (MSEs) between the inferred and true parameters for the training and the validation data as a function of the epoch for three different choices of the regularization parameter, $\lambda$. For an underestimated choice, $\lambda = 1.00 \times 10^{-6}$, the mean square error (MSE) for the validation data first decreases until epoch $\sim 1200$ and then starts increasing, even though the MSE for the training data is still decreasing. This is a symptom of over-fitting where the network overlearns the training data and fails to generalize the prediction on the unseen validation data. For an optimal choice, $\lambda = 1.69 \times 10^{-6}$, the MSEs decrease for both the training and the validation data as iteration progresses and saturate at the same value. For an overestimated choice, $\lambda = 4.00 \times 10^{-6}$, the MSEs decrease but saturate at larger values. The test data are used to quantify the accuracy of the post-training prediction of the network.

\begin{figure*}
\centering
\includegraphics[scale=0.7]{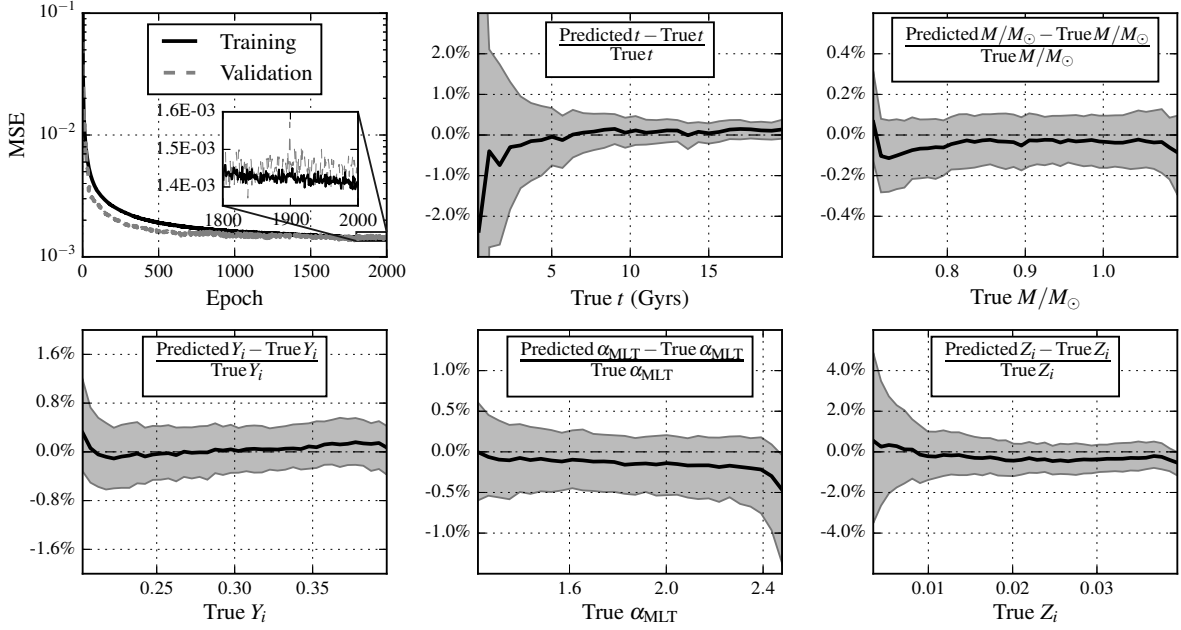}
\caption{A well-trained network should not over-fit the data, a criterion we meet, as demonstrated by the upper-left panel which shows that the training and validation data are fit to the same degree at the end of the training. The inset shows the mean square errors for the last 200 epochs. The remaining panels show errors incurred in predicting the fundamental parameters post training. The median and 1-$\sigma$ errors are plotted to illustrate the accuracy of the prediction.\label{fig4}}
\end{figure*}

\begin{figure*}
\centering
\includegraphics[scale=0.7]{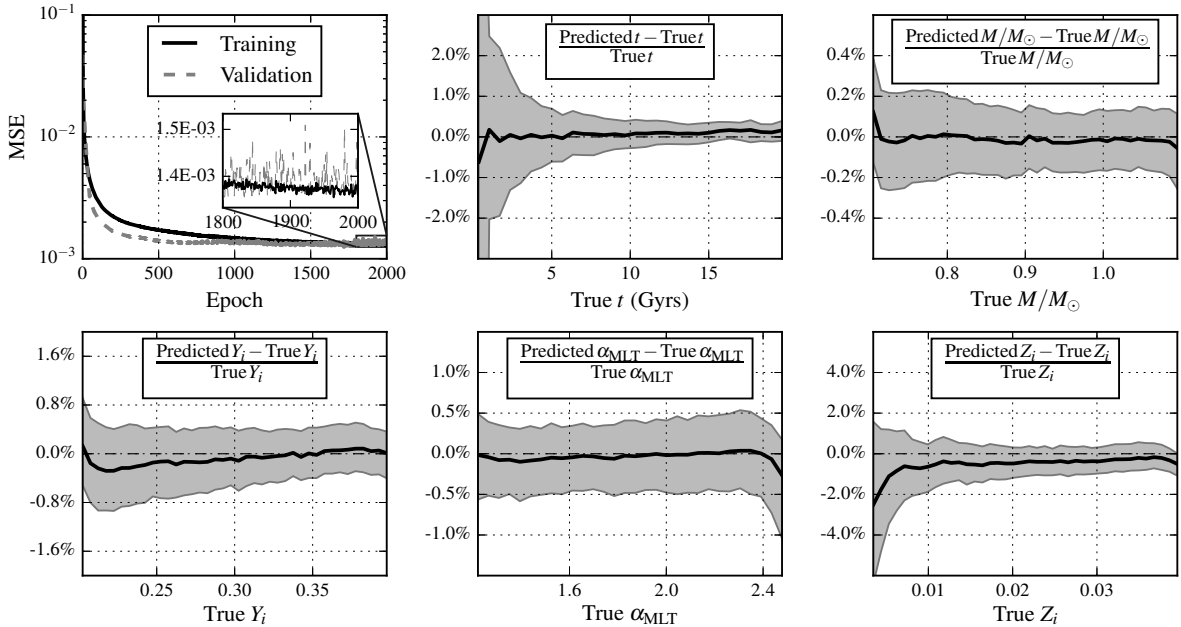}
\caption{Same as Figure~\ref{fig4} except that the corresponding network has 4 additional inputs, two $r_{02}$ at radial orders 14 and 15 and two $r_{01}$ at radial orders 15 and 16, to the current set.\label{fig5}}
\end{figure*}

The weights and biases associated with the network converge and do not over-fit the data, as shown in the upper-left panel of Figure~\ref{fig4}. The five other panels of Figure~\ref{fig4} show the performance of the network in inferring the various fundamental parameters for the test data. Methodological errors are encountered when predicting fundamental stellar parameters for the test data, depending for instance on the grid resolution of the training models. This error is known as internal error. Note from the figure that the internal errors for the mass, radius, and age are far smaller than the PLATO specifications (\citet{raue14}; masses better than 10\%, radii to 1--2\%, and age to 10\%). To test the effect of the chosen numbers and ranges of the radial orders, we train another network with four additional inputs, two $r_{02}$ at radial orders 14 and 15 and two $r_{01}$ at radial orders 15 and 16, to the current set. We find that the predictive performance of this network is not significantly different from the previous one as the MSE decreases only very slightly, from $1.39\times10^{-3}$ to $1.32\times10^{-3}$. A significantly improved predictive power would imply smaller scatter in Figure~\ref{fig5} in comparison to Figure~\ref{fig4} (not the case here). This may be due to the fact that these additional radial orders do not contain significantly new independent information. The subsequent results correspond to the network without additional radial orders.

\begin{figure*}
\centering
\includegraphics[scale=0.8]{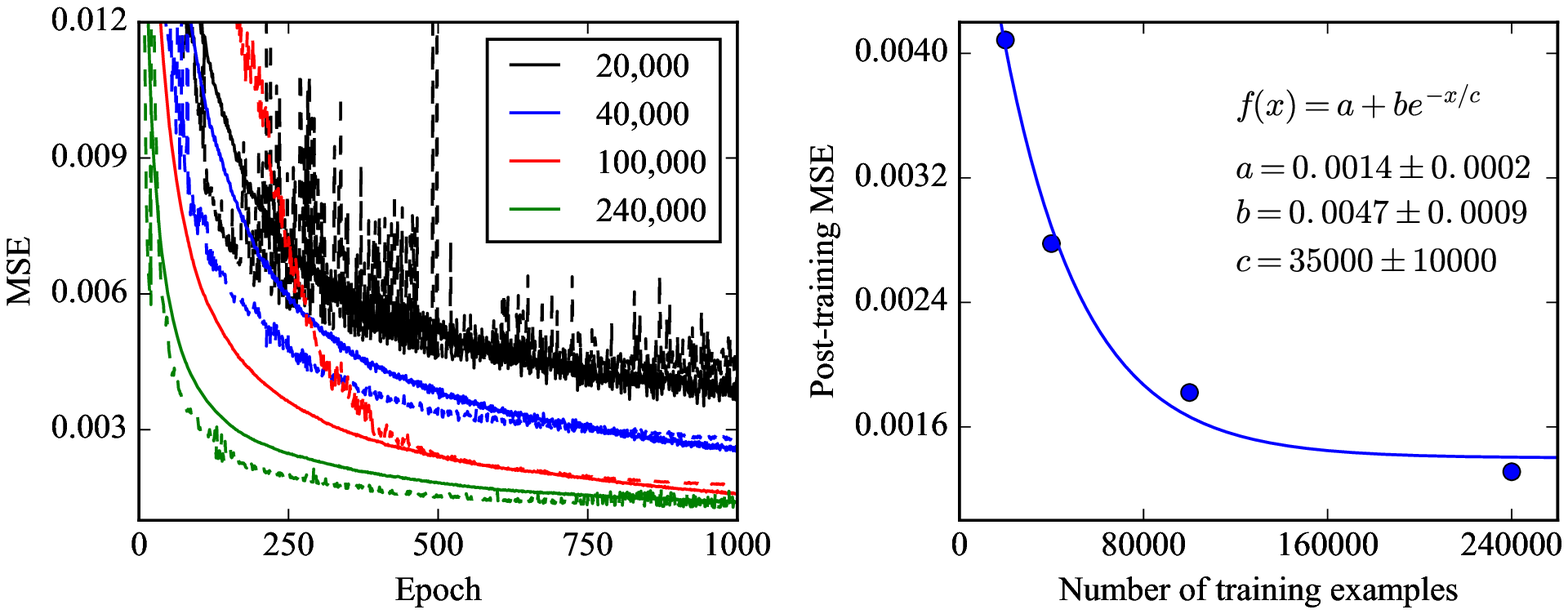}
\caption{Left panel shows the mean square errors for training (continuous lines) and validation (dashed lines) data as a function of the epoch for all the four training sets. Right panel shows the MSE for the test data at the end of the training as a function of the number of training examples.\label{fig6}}
\end{figure*}

\subsection{Test of grid resolution}
\label{subsec:grid}
To test that the above grid is dense enough for the training, we randomize the 360,000 models and take the first 120,000 for validation and testing (60,000 for each). The remaining 240,000 models are used to train four networks, with differing numbers of examples; the first with 20,000 examples, second with 40,000, third with 100,000, and the fourth with 240,000 examples. The idea is to measure the predictive power of each of these networks against the same test dataset (the 60,000 models that we have kept aside). Randomization ensures that roughly the same number of models is excluded from each evolutionary track in all cases except the last case where all the models are included. 

In Figure~\ref{fig6}, we show the MSEs for training and validation data as a function of the epoch for all the four sets in the left panel, while the right panel shows the MSE for the test data at the end of the training as a function of the number of training examples. The MSE for the test as well as the training and validation data decreases as a function of the number of training examples and saturates. To extrapolate this fall-off to infinity, we approximate it by an exponential function as shown in the right panel. This gives an estimate of the MSE at infinity (asymptotic value), which is $0.0014 \pm 0.0002$. The MSE for test data when trained with 240,000 examples is 0.0013 and is consistent with the asymptotic value. This suggests that the network performance saturates as a function of training examples before 240,000 models, and is unlikely to improve significantly if trained with denser grids. Note that the asymptotic MSE does not go to zero because the neural network only approximates the physics of this problem. However, the theoretical uncertainty introduced by the use of the network (internal error) is much smaller than the errors associated with the measurements, as described in the next section.

\subsection{Test on various stars}
\label{subsec:test}
We may now use the trained network to determine the five fundamental stellar parameters ($M$, $Y_i$, $Z_i$, $\alpha_{\rm MLT}$, and $t$). The other parameters such as radius $R$ may be determined independently. In fact, we may obtain the evolutionary parameters independent of each other by training the network to output only one parameter at a time. The networks that output all five parameters are denoted by NN5 while the one-output-parameter network is termed NN1.

We have used stellar models in the previous section to train the network. The stellar models assume a set of input physics which may differ from the physics of real stars. Moreover, the inputs of the network have observational uncertainties in case of a real star. To test the robustness of the determination of the fundamental stellar parameters against the uncertainties in the standard stellar physics and the observational uncertainties, we determine them using the trained networks (NN5 and NN1) for three well-studied real stars and one artificial star, and compare them with the known values. 

\begin{table*}
\centering
\caption{Predictions for various stars using different methods. NN5 refers to the simultaneous inference of the 5 parameters whereas NN1 infers each parameter independently. The Metc15 and Rees16 values are from \citet{metc15} and \citet{rees16}, respectively.}
\label{tab1}
\begin{tabular}{cccccccc}
\hline\hline
Star     & Method      & $M/M_\odot$               & $R/R_\odot$               & $Y_i$                     & $Z_i$                        & $t$(Gyr)               & $\alpha_{\rm MLT}$     \\
\hline
Sun      & NN5         & $0.995^{+0.058}_{-0.055}$ & ...                       & $0.291^{+0.032}_{-0.035}$ & $0.0216^{+0.0043}_{-0.0036}$ & $4.46^{+0.20}_{-0.22}$ & $1.77^{+0.11}_{-0.10}$ \\
Sun      & NN1         & $0.992^{+0.066}_{-0.059}$ & $1.001^{+0.022}_{-0.022}$ & $0.292^{+0.037}_{-0.036}$ & $0.0223^{+0.0035}_{-0.0031}$ & $4.44^{+0.18}_{-0.17}$ & $1.77^{+0.12}_{-0.11}$ \\
Sun      & ...         & $1.000$                   & $1.000$                   & $0.275$                   & $0.0200$                     & $4.60$                 & ...                    \\ 
Sun      & Grid        & $0.981 \pm 0.030$         & $0.996 \pm 0.015$         & ...                       & ...                          & $4.46 \pm 0.20$        & ...                    \\ 
16 Cyg A & NN5         & $1.085^{+0.059}_{-0.065}$ & ...                       & $0.271^{+0.051}_{-0.045}$ & $0.0257^{+0.0050}_{-0.0044}$ & $6.65^{+0.58}_{-0.67}$ & $1.86^{+0.12}_{-0.10}$ \\
16 Cyg A & NN1         & $1.083^{+0.059}_{-0.066}$ & $1.229^{+0.025}_{-0.026}$ & $0.276^{+0.037}_{-0.040}$ & $0.0283^{+0.0042}_{-0.0040}$ & $7.17^{+0.75}_{-0.79}$ & $1.87^{+0.11}_{-0.10}$ \\
16 Cyg A & Metc15      & $1.08 \pm 0.02$           & $1.229 \pm 0.008$         & $0.25 \pm 0.01$           & $0.0210 \pm 0.0020$          & $7.07 \pm 0.26$        & ...                    \\
16 Cyg A & Grid        & $1.02 \pm 0.04$           & $1.182 \pm 0.020$         & ...                       & ...                          & $5.29 \pm 1.00$        & ...                    \\
16 Cyg B & NN5         & $1.085^{+0.057}_{-0.058}$ & ...                       & $0.248^{+0.041}_{-0.042}$ & $0.0252^{+0.0044}_{-0.0037}$ & $6.69^{+0.31}_{-0.32}$ & $1.89^{+0.12}_{-0.10}$ \\
16 Cyg B & NN1         & $1.095^{+0.069}_{-0.069}$ & $1.140^{+0.023}_{-0.024}$ & $0.239^{+0.043}_{-0.041}$ & $0.0272^{+0.0032}_{-0.0031}$ & $7.14^{+0.62}_{-0.63}$ & $1.91^{+0.12}_{-0.11}$ \\
16 Cyg B & Metc15      & $1.04 \pm 0.02$           & $1.116 \pm 0.006$         & $0.26 \pm 0.01$           & $0.0220 \pm 0.0030$          & $6.74 \pm 0.24$        & ...                    \\
16 Cyg B & Grid        & $1.08 \pm 0.03$           & $1.124 \pm 0.015$         & ...                       & ...                          & $6.10 \pm 0.50$        & ...                    \\
Elvis    & NN5         & $0.967^{+0.086}_{-0.073}$ & ...                       & $0.291^{+0.055}_{-0.061}$ & $0.0189^{+0.0041}_{-0.0031}$ & $6.23^{+0.64}_{-0.64}$ & $1.84^{+0.13}_{-0.11}$ \\
Elvis    & NN1         & $0.986^{+0.094}_{-0.084}$ & $1.069^{+0.033}_{-0.032}$ & $0.259^{+0.068}_{-0.071}$ & $0.0238^{+0.0043}_{-0.0040}$ & $7.14^{+0.97}_{-1.01}$ & $1.89^{+0.16}_{-0.13}$ \\
Elvis    & Rees16      & $1.000$                   & $1.087$                   & $0.267$                   & $0.0176$                     & $6.84$                 & $1.67$                 \\
Elvis    & Grid        & $1.037 \pm 0.03$          & $1.103 \pm 0.015$         & ...                       & ...                          & $6.07 \pm 0.40$        & ...                    \\ 
\hline
\end{tabular}
\end{table*}

\subsubsection{The Sun}
We use seismic data from \citet{chap07} and spectroscopic measurements from \citet{metc15} to compute the required inputs for the network. We generate 40,000 realizations of the input, using its covariance matrix and assuming that the errors on the input follow a multivariate Gaussian distribution, to obtain the distribution of the output parameters. We compute the median for the parameter values and $16^{\rm th}$ and $84^{\rm th}$ percentiles for negative and positive uncertainties on them. The internal error, which was quantified in the previous section, contributes to the uncertainty. We note that it is typically difficult in conventional methods to disentangle the internal error and the error due to the uncertainties in the measurements.  

\begin{figure*}
\centering
\includegraphics[scale=0.7]{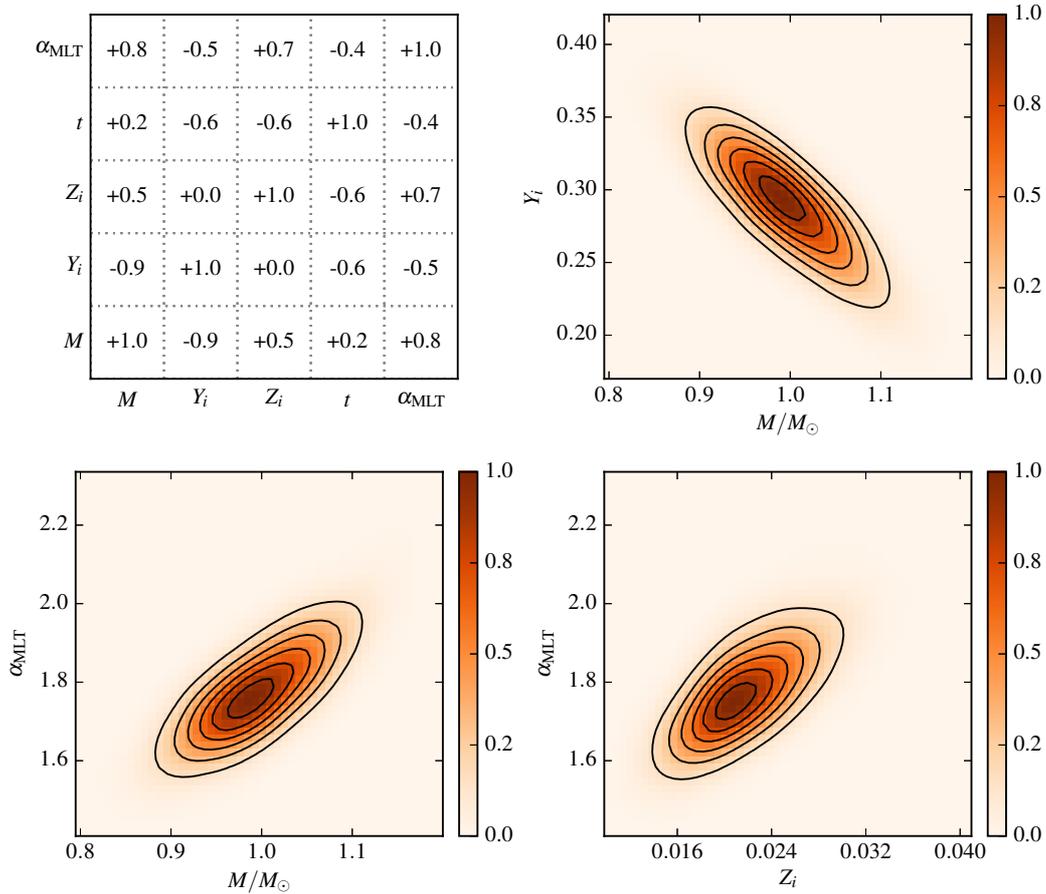}
\caption{Tradeoffs between the inferred parameters from a Monte-Carlo simulation with 40,000 realizations of the solar data. The full correlation matrix is shown on the upper-left panel, whereas the other panels show the tradeoffs between specific parameters. Indeed, $Y_i$ is strongly anti-correlated with $M$. More interestingly, the network reveals that the pairs $(\alpha_{\rm MLT},M)$ and $(\alpha_{\rm MLT},Z_i)$ are correlated. The ideal inference would correspond to a situation when all the parameters are independent of each other (and not the case here).\label{fig7}}
\end{figure*}

Table~\ref{tab1} lists the parameters for Sun obtained using NN5 and NN1. We retrieve the solar parameters accurately, to within  5.5\%, 2.2\%, and 4.5\% of the mass, radius, and the age of the Sun, respectively. Although the precision is well within the PLATO specifications, it is poor given the precision of the solar data. This may in part be attributed to the limited number of seismic inputs used (we used small number of seismic inputs anticipating the fact that a small set of modes is observed in distant stars). More importantly, correlations among the stellar parameters, for instance the anti-correlation between $M$ and $Y_i$ \citep[see, e.g.,][]{metc09,lebr14}, degrade the quality of the inference given that the measurements have associated uncertainties. The top-left panel of Figure~\ref{fig7} shows the correlation matrix of the stellar parameters which is obtained from the distribution of the output parameters. The other panels in the figure show the distribution as a function of two stellar parameters. It is seen that $M$ and $Y_i$ are anti-correlated while $M$ and $\alpha_{\rm MLT}$ show a positive correlation. While fixing some stellar parameters, e.g. $Y_i$, using the Helium-to-metal enrichment law and $\alpha_{\rm MLT}$ using solar calibration, would lead to higher precision in the determination of the other parameters, they also become prone to large systematic errors (e.g., \citet{silv15} shows that the systematics associated with the fixed $Y_i$ is comparable to the statistical uncertainties). The analysis of the acoustic glitches provides constraint on $Y_i$ \citep[see, e.g.,][]{basu04,mont05,houd07,verm14a}, and can be used to improve on the precision.

\subsubsection{16 Cyg A \& B and Elvis}
16 Cyg A \& B form a binary system and are among the brightest stars observed by {\it Kepler}. This system has been central to asteroseismic studies in the recent past \citep[see, e.g.,][]{metc12,grub13,verm14a,metc15,buld16}. We use the same spectroscopic and seismic data for the 16 Cyg A \& B as used in \citet{metc15} for the input of the network. Elvis is an artificial star constructed for a hare-and-hounds exercise \citep{rees16} using the CLES stellar evolution code \citep{scuf08}. We choose Elvis out of the 10 artificial stars studied in \citet{rees16} because this is the only star whose mass is in the range [0.7--1.1]$M_\odot$ and has a sufficient number of listed oscillation frequencies required as inputs to our network. In contrast to the input physics used in our training models, \citet{rees16} used metallicity mixtures from \citet{gn93} and did not incorporate the diffusion of Helium and heavy elements for Elvis. 

The fundamental parameters obtained using NN5 and NN1 are listed in Table~\ref{tab1}. The values from NN5, NN1, Metc15 \citep{metc15}, and Rees16 \citep{rees16} are in agreement to within the error bars. The average precisions on mass, radius, and age are around 6.4\%, 2.3\%, and 8.1\%, respectively. The measurement precision of the PLATO targets is expected to be similar to these stars, and therefore, it should be possible to infer the stellar parameters to specified precisions using this method.

\subsection{Comparison with a conventional method}
\label{subsec:comp}
To compare the performance of this method with a grid based technique, we apply a method described in the section~5.1 of \citet{appo15} to the grid constructed for the training, and estimate the masses, radii, and the ages of all the four stars. The values are listed in the fourth row of Table~\ref{tab1} for every star. The relatively inaccurate parameter values suggest that the grid is too sparse for this conventional method. However, the neural network appears to extract enough information from the same grid to make accurate prediction. This shows the learning capability of the neural networks. 

\section{Summary}
\label{sec:sum}
The lack of a formal inverse relationship between the evolutionary parameters and stellar observations has led us to create an effective empirical equivalent. In the current context, the neural network can be thought of as a high-dimensional look-up table, or a concise map of the space in which all the inputs and outputs lie. We note the following about our network:
\begin{itemize}
\item it is optimally trained for the parameter regime that we have studied, and is not expected to improve when trained on more data. This is demonstrated by the fact that the mean square error nearly falls to its asymptotic value when trained with 360,000 models.
\item it is computationally cheap ($\sim 50$ cpu hours) to retrain the network, useful when changing the inputs or outputs.
\item the predictive power of the network does not change appreciably upon increasing the number of radial orders of frequency measurements used as inputs. 
\item preliminary indications are that the network is robust to changes in the input physics, as suggested by its accurate inferences of the parameters of an artificial star (Elvis), whose structure was computed using a different prescription, and the parameters of three real stars (the Sun, 16 Cyg A and B). 
\item it can be used to determine stellar parameters with precision similar to detailed grid modeling methods, but takes orders of magnitude less time.
\end{itemize}
The success of neural networks in the current application encourages a more thorough examination of its capabilities.

\section*{Acknowledgements}
We thank the anonymous referee for comments which have improved the quality of the manuscript. SH acknowledges support from Ramanujan fellowship SB/S2/RJN-73, the Max-Planck partner group program, and the Centre for Space Science at the New York University, Abu Dhabi. 


\begin{thebibliography}{}
\makeatletter
\relax
\def\mn@urlcharsother{\let\do\@makeother \do\$\do\&\do\#\do\^\do\_\do\%\do\~}
\def\mn@doi{\begingroup\mn@urlcharsother \@ifnextchar [ {\mn@doi@}
  {\mn@doi@[]}}
\def\mn@doi@[#1]#2{\def\@tempa{#1}\ifx\@tempa\@empty \href
  {http://dx.doi.org/#2} {doi:#2}\else \href {http://dx.doi.org/#2} {#1}\fi
  \endgroup}
\def\mn@eprint#1#2{\mn@eprint@#1:#2::\@nil}
\def\mn@eprint@arXiv#1{\href {http://arxiv.org/abs/#1} {{\tt arXiv:#1}}}
\def\mn@eprint@dblp#1{\href {http://dblp.uni-trier.de/rec/bibtex/#1.xml}
  {dblp:#1}}
\def\mn@eprint@#1:#2:#3:#4\@nil{\def\@tempa {#1}\def\@tempb {#2}\def\@tempc
  {#3}\ifx \@tempc \@empty \let \@tempc \@tempb \let \@tempb \@tempa \fi \ifx
  \@tempb \@empty \def\@tempb {arXiv}\fi \@ifundefined
  {mn@eprint@\@tempb}{\@tempb:\@tempc}{\expandafter \expandafter \csname
  mn@eprint@\@tempb\endcsname \expandafter{\@tempc}}}

\bibitem[\protect\citeauthoryear{{Aerts}, {Christensen-Dalsgaard}  \&
  {Kurtz}}{{Aerts} et~al.}{2010}]{aert10}
{Aerts} C.,  {Christensen-Dalsgaard} J.,   {Kurtz} D.~W.,  2010,
  {Asteroseismology}.
Springer-Verlag, Heidelberg

\bibitem[\protect\citeauthoryear{{Angulo} et~al.,}{{Angulo}
  et~al.}{1999}]{angu99}
{Angulo} C.,  et~al., 1999, \mn@doi [Nuclear Physics A]
  {10.1016/S0375-9474(99)00030-5}, \href
  {http://adsabs.harvard.edu/abs/1999NuPhA.656....3A} {656, 3}

\bibitem[\protect\citeauthoryear{{Appourchaux} et~al.,}{{Appourchaux}
  et~al.}{2015}]{appo15}
{Appourchaux} T.,  et~al., 2015, \mn@doi [\aap] {10.1051/0004-6361/201526610},
  \href {http://adsabs.harvard.edu/abs/2015A%26A...582A..25A} {582, A25}

\bibitem[\protect\citeauthoryear{{Badnell}, {Bautista}, {Butler}, {Delahaye},
  {Mendoza}, {Palmeri}, {Zeippen}  \& {Seaton}}{{Badnell}
  et~al.}{2005}]{badn05}
{Badnell} N.~R.,  {Bautista} M.~A.,  {Butler} K.,  {Delahaye} F.,  {Mendoza}
  C.,  {Palmeri} P.,  {Zeippen} C.~J.,   {Seaton} M.~J.,  2005, \mn@doi
  [\mnras] {10.1111/j.1365-2966.2005.08991.x}, \href
  {http://adsabs.harvard.edu/abs/2005MNRAS.360..458B} {360, 458}

\bibitem[\protect\citeauthoryear{{Baglin}, {Auvergne}, {Barge}, {Deleuil},
  {Michel}  \& {CoRoT Exoplanet Science Team}}{{Baglin} et~al.}{2009}]{bagl09}
{Baglin} A.,  {Auvergne} M.,  {Barge} P.,  {Deleuil} M.,  {Michel} E.,   {CoRoT
  Exoplanet Science Team} 2009, in {Pont} F.,  {Sasselov} D.,   {Holman} M.~J.,
   eds,  IAU Symposium Vol. 253, IAU Symposium. pp 71--81,
  \mn@doi{10.1017/S1743921308026252}

\bibitem[\protect\citeauthoryear{Bastien, Lamblin, Pascanu, Bergstra,
  Goodfellow, Bergeron, Bouchard  \& Bengio}{Bastien et~al.}{2012}]{bast12}
Bastien F.,  Lamblin P.,  Pascanu R.,  Bergstra J.,  Goodfellow I.~J.,
  Bergeron A.,  Bouchard N.,   Bengio Y.,  2012, Theano: new features and speed
  improvements, Deep Learning and Unsupervised Feature Learning NIPS 2012
  Workshop

\bibitem[\protect\citeauthoryear{{Basu}, {Mazumdar}, {Antia}  \&
  {Demarque}}{{Basu} et~al.}{2004}]{basu04}
{Basu} S.,  {Mazumdar} A.,  {Antia} H.~M.,   {Demarque} P.,  2004, \mn@doi
  [\mnras] {10.1111/j.1365-2966.2004.07644.x}, \href
  {http://adsabs.harvard.edu/abs/2004MNRAS.350..277B} {350, 277}

\bibitem[\protect\citeauthoryear{Bergstra \& Bengio}{Bergstra \&
  Bengio}{2012}]{berg12}
Bergstra J.,  Bengio Y.,  2012, Journal of Machine Learning Research, 13, 281

\bibitem[\protect\citeauthoryear{Bergstra et~al.,}{Bergstra
  et~al.}{2010}]{berg10}
Bergstra J.,  et~al., 2010, in Proceedings of the Python for Scientific
  Computing Conference ({SciPy}).

\bibitem[\protect\citeauthoryear{{Borucki} et~al.,}{{Borucki}
  et~al.}{2009}]{boru09}
{Borucki} W.,  et~al., 2009, in {Pont} F.,  {Sasselov} D.,   {Holman} M.~J.,
  eds,  IAU Symposium Vol. 253, IAU Symposium. pp 289--299,
  \mn@doi{10.1017/S1743921308026513}

\bibitem[\protect\citeauthoryear{Bottou}{Bottou}{2010}]{bott10}
Bottou L.,  2010, in Lechevallier Y.,  Saporta G.,  eds, Proceedings of the
  19th International Conference on Computational Statistics (COMPSTAT'2010).
  Springer, Paris, France, pp 177--187, \url
  {http://leon.bottou.org/papers/bottou-2010}

\bibitem[\protect\citeauthoryear{{Brown} \& {Gilliland}}{{Brown} \&
  {Gilliland}}{1994}]{brow94}
{Brown} T.~M.,  {Gilliland} R.~L.,  1994, \mn@doi [\araa]
  {10.1146/annurev.aa.32.090194.000345}, \href
  {http://adsabs.harvard.edu/abs/1994ARA%26A..32...37B} {32, 37}

\bibitem[\protect\citeauthoryear{{Buldgen}, {Reese}  \& {Dupret}}{{Buldgen}
  et~al.}{2016}]{buld16}
{Buldgen} G.,  {Reese} D.~R.,   {Dupret} M.~A.,  2016, \mn@doi [\aap]
  {10.1051/0004-6361/201527032}, \href
  {http://adsabs.harvard.edu/abs/2016A%26A...585A.109B} {585, A109}

\bibitem[\protect\citeauthoryear{{Casagrande} et~al.,}{{Casagrande}
  et~al.}{2014}]{casa14}
{Casagrande} L.,  et~al., 2014, \mn@doi [\apj] {10.1088/0004-637X/787/2/110},
  \href {http://adsabs.harvard.edu/abs/2014ApJ...787..110C} {787, 110}

\bibitem[\protect\citeauthoryear{{Chaplin}, {Elsworth}, {Miller}, {Verner}  \&
  {New}}{{Chaplin} et~al.}{2007}]{chap07}
{Chaplin} W.~J.,  {Elsworth} Y.,  {Miller} B.~A.,  {Verner} G.~A.,   {New} R.,
  2007, \mn@doi [\apj] {10.1086/512543}, \href
  {http://adsabs.harvard.edu/abs/2007ApJ...659.1749C} {659, 1749}

\bibitem[\protect\citeauthoryear{{Chaplin} et~al.,}{{Chaplin}
  et~al.}{2011}]{chap11}
{Chaplin} W.~J.,  et~al., 2011, \mn@doi [Science] {10.1126/science.1201827},
  \href {http://adsabs.harvard.edu/abs/2011Sci...332..213C} {332, 213}

\bibitem[\protect\citeauthoryear{{Chaplin} et~al.,}{{Chaplin}
  et~al.}{2014}]{chap14}
{Chaplin} W.~J.,  et~al., 2014, \mn@doi [\apjs] {10.1088/0067-0049/210/1/1},
  \href {http://adsabs.harvard.edu/abs/2014ApJS..210....1C} {210, 1}

\bibitem[\protect\citeauthoryear{{Christensen-Dalsgaard}}{{Christensen-Dalsgaard}}{1993}]{jcd93}
{Christensen-Dalsgaard} J.,  1993, in {Brown} T.~M.,  ed.,  Astronomical
  Society of the Pacific Conference Series Vol. 42, GONG 1992. Seismic
  Investigation of the Sun and Stars. p.~347

\bibitem[\protect\citeauthoryear{{Christensen-Dalsgaard}}{{Christensen-Dalsgaard}}{2004}]{jcd04}
{Christensen-Dalsgaard} J.,  2004, \mn@doi [\solphys]
  {10.1023/B:SOLA.0000031392.43227.7d}, \href
  {http://adsabs.harvard.edu/abs/2004SoPh..220..137C} {220, 137}

\bibitem[\protect\citeauthoryear{{Christensen-Dalsgaard}}{{Christensen-Dalsgaard}}{2008}]{jcd08}
{Christensen-Dalsgaard} J.,  2008, \mn@doi [\apss] {10.1007/s10509-007-9689-z},
  \href {http://adsabs.harvard.edu/abs/2008Ap%26SS.316..113C} {316, 113}

\bibitem[\protect\citeauthoryear{{Christensen-Dalsgaard} \&
  {Frandsen}}{{Christensen-Dalsgaard} \& {Frandsen}}{1988}]{jcd88}
{Christensen-Dalsgaard} J.,  {Frandsen} S.,  eds, 1988, {Advances in helio- and
  asteroseismology; Proceedings of the Symposium, Aarhus, Denmark, July 7-11,
  1986}  IAU Symposium Vol. 123

\bibitem[\protect\citeauthoryear{{Christensen-Dalsgaard} \&
  {Thompson}}{{Christensen-Dalsgaard} \& {Thompson}}{1997}]{jcd97}
{Christensen-Dalsgaard} J.,  {Thompson} M.~J.,  1997, \mn@doi [\mnras]
  {10.1093/mnras/284.3.527}, \href
  {http://adsabs.harvard.edu/abs/1997MNRAS.284..527C} {284, 527}

\bibitem[\protect\citeauthoryear{{Christensen-Dalsgaard}, {Dappen}  \&
  {Lebreton}}{{Christensen-Dalsgaard} et~al.}{1988}]{jcd88b}
{Christensen-Dalsgaard} J.,  {Dappen} W.,   {Lebreton} Y.,  1988, \mn@doi
  [\nat] {10.1038/336634a0}, \href
  {http://adsabs.harvard.edu/abs/1988Natur.336..634C} {336, 634}

\bibitem[\protect\citeauthoryear{{Christensen-Dalsgaard}
  et~al.,}{{Christensen-Dalsgaard} et~al.}{1996}]{jcd96}
{Christensen-Dalsgaard} J.,  et~al., 1996, \mn@doi [Science]
  {10.1126/science.272.5266.1286}, \href
  {http://adsabs.harvard.edu/abs/1996Sci...272.1286C} {272, 1286}

\bibitem[\protect\citeauthoryear{Cox \& Giuli}{Cox \& Giuli}{1968}]{cox68}
Cox J.,  Giuli R.,  1968, Principles of Stellar Structure: Physical principles.
No.~v. 1 in Principles of Stellar Structure, Gordon and Breach, \url
  {http://books.google.co.in/books?id=TdhEAAAAIAAJ}

\bibitem[\protect\citeauthoryear{{Dziembowski}, {Paterno}  \&
  {Ventura}}{{Dziembowski} et~al.}{1988}]{dzie88}
{Dziembowski} W.~A.,  {Paterno} L.,   {Ventura} R.,  1988, \aap, \href
  {http://adsabs.harvard.edu/abs/1988A%26A...200..213D} {200, 213}

\bibitem[\protect\citeauthoryear{{Epstein} et~al.,}{{Epstein}
  et~al.}{2014}]{epst14}
{Epstein} C.~R.,  et~al., 2014, \mn@doi [\apjl] {10.1088/2041-8205/785/2/L28},
  \href {http://adsabs.harvard.edu/abs/2014ApJ...785L..28E} {785, L28}

\bibitem[\protect\citeauthoryear{{Escobar} et~al.,}{{Escobar}
  et~al.}{2012}]{esco12}
{Escobar} M.~E.,  et~al., 2012, \mn@doi [\aap] {10.1051/0004-6361/201218969},
  \href {http://adsabs.harvard.edu/abs/2012A%26A...543A..96E} {543, A96}

\bibitem[\protect\citeauthoryear{{Ferguson}, {Alexander}, {Allard}, {Barman},
  {Bodnarik}, {Hauschildt}, {Heffner-Wong}  \& {Tamanai}}{{Ferguson}
  et~al.}{2005}]{ferg05}
{Ferguson} J.~W.,  {Alexander} D.~R.,  {Allard} F.,  {Barman} T.,  {Bodnarik}
  J.~G.,  {Hauschildt} P.~H.,  {Heffner-Wong} A.,   {Tamanai} A.,  2005,
  \mn@doi [\apj] {10.1086/428642}, \href
  {http://adsabs.harvard.edu/abs/2005ApJ...623..585F} {623, 585}

\bibitem[\protect\citeauthoryear{{Gilliland}, {McCullough}, {Nelan}, {Brown},
  {Charbonneau}, {Nutzman}, {Christensen-Dalsgaard}  \& {Kjeldsen}}{{Gilliland}
  et~al.}{2011}]{gill11}
{Gilliland} R.~L.,  {McCullough} P.~R.,  {Nelan} E.~P.,  {Brown} T.~M.,
  {Charbonneau} D.,  {Nutzman} P.,  {Christensen-Dalsgaard} J.,   {Kjeldsen}
  H.,  2011, \mn@doi [\apj] {10.1088/0004-637X/726/1/2}, \href
  {http://adsabs.harvard.edu/abs/2011ApJ...726....2G} {726, 2}

\bibitem[\protect\citeauthoryear{{Gilliland} et~al.,}{{Gilliland}
  et~al.}{2013}]{gill13}
{Gilliland} R.~L.,  et~al., 2013, \mn@doi [\apj] {10.1088/0004-637X/766/1/40},
  \href {http://adsabs.harvard.edu/abs/2013ApJ...766...40G} {766, 40}

\bibitem[\protect\citeauthoryear{Glorot, Bordes  \& Bengio}{Glorot
  et~al.}{2011}]{glor11}
Glorot X.,  Bordes A.,   Bengio Y.,  2011, in Proceedings of the Fourteenth
  International Conference on Artificial Intelligence and Statistics, {AISTATS}
  2011, Fort Lauderdale, USA, April 11-13, 2011. pp 315--323, \url
  {http://www.jmlr.org/proceedings/papers/v15/glorot11a/glorot11a.pdf}

\bibitem[\protect\citeauthoryear{{Grevesse} \& {Noels}}{{Grevesse} \&
  {Noels}}{1993}]{gn93}
{Grevesse} N.,  {Noels} A.,  1993, in {Prantzos} N.,  {Vangioni-Flam} E.,
  {Casse} M.,  eds, Origin and Evolution of the Elements. pp 15--25

\bibitem[\protect\citeauthoryear{{Grevesse} \& {Sauval}}{{Grevesse} \&
  {Sauval}}{1998}]{gs98}
{Grevesse} N.,  {Sauval} A.~J.,  1998, \mn@doi [\ssr]
  {10.1023/A:1005161325181}, \href
  {http://adsabs.harvard.edu/abs/1998SSRv...85..161G} {85, 161}

\bibitem[\protect\citeauthoryear{{Gruberbauer}, {Guenther}  \&
  {Kallinger}}{{Gruberbauer} et~al.}{2012}]{grub12}
{Gruberbauer} M.,  {Guenther} D.~B.,   {Kallinger} T.,  2012, \mn@doi [\apj]
  {10.1088/0004-637X/749/2/109}, \href
  {http://adsabs.harvard.edu/abs/2012ApJ...749..109G} {749, 109}

\bibitem[\protect\citeauthoryear{{Gruberbauer}, {Guenther}, {MacLeod}  \&
  {Kallinger}}{{Gruberbauer} et~al.}{2013}]{grub13}
{Gruberbauer} M.,  {Guenther} D.~B.,  {MacLeod} K.,   {Kallinger} T.,  2013,
  \mn@doi [\mnras] {10.1093/mnras/stt1289}, \href
  {http://adsabs.harvard.edu/abs/2013MNRAS.435..242G} {435, 242}

\bibitem[\protect\citeauthoryear{{Houdek} \& {Gough}}{{Houdek} \&
  {Gough}}{2007}]{houd07}
{Houdek} G.,  {Gough} D.~O.,  2007, \mn@doi [\mnras]
  {10.1111/j.1365-2966.2006.11325.x}, \href
  {http://adsabs.harvard.edu/abs/2007MNRAS.375..861H} {375, 861}

\bibitem[\protect\citeauthoryear{{Huber} et~al.,}{{Huber}
  et~al.}{2011}]{hube11}
{Huber} D.,  et~al., 2011, \mn@doi [\apj] {10.1088/0004-637X/743/2/143}, \href
  {http://adsabs.harvard.edu/abs/2011ApJ...743..143H} {743, 143}

\bibitem[\protect\citeauthoryear{{Huber} et~al.,}{{Huber}
  et~al.}{2012}]{hube12}
{Huber} D.,  et~al., 2012, \mn@doi [\apj] {10.1088/0004-637X/760/1/32}, \href
  {http://adsabs.harvard.edu/abs/2012ApJ...760...32H} {760, 32}

\bibitem[\protect\citeauthoryear{{Huber} et~al.,}{{Huber}
  et~al.}{2013}]{hube13}
{Huber} D.,  et~al., 2013, \mn@doi [\apj] {10.1088/0004-637X/767/2/127}, \href
  {http://adsabs.harvard.edu/abs/2013ApJ...767..127H} {767, 127}

\bibitem[\protect\citeauthoryear{{Imbriani} et~al.,}{{Imbriani}
  et~al.}{2005}]{imbr05}
{Imbriani} G.,  et~al., 2005, \mn@doi [European Physical Journal A]
  {10.1140/epja/i2005-10138-7}, \href
  {http://adsabs.harvard.edu/abs/2005EPJA...25..455I} {25, 455}

\bibitem[\protect\citeauthoryear{{Kjeldsen} \& {Bedding}}{{Kjeldsen} \&
  {Bedding}}{1995}]{kjel95}
{Kjeldsen} H.,  {Bedding} T.~R.,  1995, \aap, \href
  {http://adsabs.harvard.edu/abs/1995A%26A...293...87K} {293, 87}

\bibitem[\protect\citeauthoryear{{Kunz}, {Fey}, {Jaeger}, {Mayer}, {Hammer},
  {Staudt}, {Harissopulos}  \& {Paradellis}}{{Kunz} et~al.}{2002}]{kunz02}
{Kunz} R.,  {Fey} M.,  {Jaeger} M.,  {Mayer} A.,  {Hammer} J.~W.,  {Staudt} G.,
   {Harissopulos} S.,   {Paradellis} T.,  2002, \mn@doi [\apj]
  {10.1086/338384}, \href {http://adsabs.harvard.edu/abs/2002ApJ...567..643K}
  {567, 643}

\bibitem[\protect\citeauthoryear{LeCun, Bottou, Orr  \& M{\"u}ller}{LeCun
  et~al.}{1998}]{lecu98}
LeCun Y.,  Bottou L.,  Orr G.~B.,   M{\"u}ller K.~R.,  1998, Neural Networks:
  Tricks of the Trade.
Springer Berlin Heidelberg, Berlin, Heidelberg, pp 9--50,
  \mn@doi{10.1007/3-540-49430-8_2}, \url
  {http://dx.doi.org/10.1007/3-540-49430-8_2}

\bibitem[\protect\citeauthoryear{LeCun, Bengio  \& Hinton}{LeCun
  et~al.}{2015}]{lecu15}
LeCun Y.,  Bengio Y.,   Hinton G.,  2015, Nature, 521, 436

\bibitem[\protect\citeauthoryear{{Lebreton} \& {Goupil}}{{Lebreton} \&
  {Goupil}}{2014}]{lebr14}
{Lebreton} Y.,  {Goupil} M.~J.,  2014, \mn@doi [\aap]
  {10.1051/0004-6361/201423797}, \href
  {http://adsabs.harvard.edu/abs/2014A%26A...569A..21L} {569, A21}

\bibitem[\protect\citeauthoryear{{Liu}, {Bi}, {Li}, {Liu}, {Tian}  \&
  {Ge}}{{Liu} et~al.}{2014}]{liu14}
{Liu} K.,  {Bi} S.-L.,  {Li} T.-D.,  {Liu} Z.-E.,  {Tian} Z.-J.,   {Ge} Z.-S.,
  2014, \mn@doi [Research in Astronomy and Astrophysics]
  {10.1088/1674-4527/14/11/008}, \href
  {http://adsabs.harvard.edu/abs/2014RAA....14.1447L} {14, 1447}

\bibitem[\protect\citeauthoryear{{Mathur} et~al.,}{{Mathur}
  et~al.}{2012}]{math12}
{Mathur} S.,  et~al., 2012, \mn@doi [\apj] {10.1088/0004-637X/749/2/152}, \href
  {http://adsabs.harvard.edu/abs/2012ApJ...749..152M} {749, 152}

\bibitem[\protect\citeauthoryear{{Metcalfe}, {Creevey}  \&
  {Christensen-Dalsgaard}}{{Metcalfe} et~al.}{2009}]{metc09}
{Metcalfe} T.~S.,  {Creevey} O.~L.,   {Christensen-Dalsgaard} J.,  2009,
  \mn@doi [\apj] {10.1088/0004-637X/699/1/373}, \href
  {http://adsabs.harvard.edu/abs/2009ApJ...699..373M} {699, 373}

\bibitem[\protect\citeauthoryear{{Metcalfe} et~al.,}{{Metcalfe}
  et~al.}{2012}]{metc12}
{Metcalfe} T.~S.,  et~al., 2012, \mn@doi [\apjl] {10.1088/2041-8205/748/1/L10},
  \href {http://adsabs.harvard.edu/abs/2012ApJ...748L..10M} {748, L10}

\bibitem[\protect\citeauthoryear{{Metcalfe} et~al.,}{{Metcalfe}
  et~al.}{2014}]{metc14}
{Metcalfe} T.~S.,  et~al., 2014, \mn@doi [\apjs] {10.1088/0067-0049/214/2/27},
  \href {http://adsabs.harvard.edu/abs/2014ApJS..214...27M} {214, 27}

\bibitem[\protect\citeauthoryear{{Metcalfe}, {Creevey}  \& {Davies}}{{Metcalfe}
  et~al.}{2015}]{metc15}
{Metcalfe} T.~S.,  {Creevey} O.~L.,   {Davies} G.~R.,  2015, \mn@doi [\apjl]
  {10.1088/2041-8205/811/2/L37}, \href
  {http://adsabs.harvard.edu/abs/2015ApJ...811L..37M} {811, L37}

\bibitem[\protect\citeauthoryear{{Miglio} et~al.,}{{Miglio}
  et~al.}{2013}]{migl13}
{Miglio} A.,  et~al., 2013, \mn@doi [\mnras] {10.1093/mnras/sts345}, \href
  {http://adsabs.harvard.edu/abs/2013MNRAS.429..423M} {429, 423}

\bibitem[\protect\citeauthoryear{{Monteiro} \& {Thompson}}{{Monteiro} \&
  {Thompson}}{2005}]{mont05}
{Monteiro} M.~J.~P.~F.~G.,  {Thompson} M.~J.,  2005, \mn@doi [\mnras]
  {10.1111/j.1365-2966.2005.09246.x}, \href
  {http://adsabs.harvard.edu/abs/2005MNRAS.361.1187M} {361, 1187}

\bibitem[\protect\citeauthoryear{{Nutzman} et~al.,}{{Nutzman}
  et~al.}{2011}]{nutz11}
{Nutzman} P.,  et~al., 2011, \mn@doi [\apj] {10.1088/0004-637X/726/1/3}, \href
  {http://adsabs.harvard.edu/abs/2011ApJ...726....3N} {726, 3}

\bibitem[\protect\citeauthoryear{{Ot{\'{\i}} Floranes}, {Christensen-Dalsgaard}
   \& {Thompson}}{{Ot{\'{\i}} Floranes} et~al.}{2005}]{oti05}
{Ot{\'{\i}} Floranes} H.,  {Christensen-Dalsgaard} J.,   {Thompson} M.~J.,
  2005, \mn@doi [\mnras] {10.1111/j.1365-2966.2004.08487.x}, \href
  {http://adsabs.harvard.edu/abs/2005MNRAS.356..671O} {356, 671}

\bibitem[\protect\citeauthoryear{{Paxton}, {Bildsten}, {Dotter}, {Herwig},
  {Lesaffre}  \& {Timmes}}{{Paxton} et~al.}{2011}]{paxt11}
{Paxton} B.,  {Bildsten} L.,  {Dotter} A.,  {Herwig} F.,  {Lesaffre} P.,
  {Timmes} F.,  2011, \mn@doi [\apjs] {10.1088/0067-0049/192/1/3}, \href
  {http://adsabs.harvard.edu/abs/2011ApJS..192....3P} {192, 3}

\bibitem[\protect\citeauthoryear{{Paxton} et~al.,}{{Paxton}
  et~al.}{2013}]{paxt13}
{Paxton} B.,  et~al., 2013, \mn@doi [\apjs] {10.1088/0067-0049/208/1/4}, \href
  {http://adsabs.harvard.edu/abs/2013ApJS..208....4P} {208, 4}

\bibitem[\protect\citeauthoryear{{Rauer} et~al.,}{{Rauer}
  et~al.}{2014}]{raue14}
{Rauer} H.,  et~al., 2014, \mn@doi [Experimental Astronomy]
  {10.1007/s10686-014-9383-4}, \href
  {http://adsabs.harvard.edu/abs/2014ExA....38..249R} {38, 249}

\bibitem[\protect\citeauthoryear{{Reese} et~al.,}{{Reese}
  et~al.}{2016}]{rees16}
{Reese} D.~R.,  et~al., 2016, preprint, \href
  {http://adsabs.harvard.edu/abs/2016arXiv160408404R} {} (\mn@eprint {arXiv}
  {1604.08404})

\bibitem[\protect\citeauthoryear{{Ricker} et~al.,}{{Ricker}
  et~al.}{2014}]{rick14}
{Ricker} G.~R.,  et~al., 2014, in Society of Photo-Optical Instrumentation
  Engineers (SPIE) Conference Series. p. 914320 (\mn@eprint {arXiv}
  {1406.0151}), \mn@doi{10.1117/12.2063489}

\bibitem[\protect\citeauthoryear{{Rogers} \& {Nayfonov}}{{Rogers} \&
  {Nayfonov}}{2002}]{roge02}
{Rogers} F.~J.,  {Nayfonov} A.,  2002, \mn@doi [\apj] {10.1086/341894}, \href
  {http://adsabs.harvard.edu/abs/2002ApJ...576.1064R} {576, 1064}

\bibitem[\protect\citeauthoryear{{Roxburgh}}{{Roxburgh}}{2005}]{roxb05}
{Roxburgh} I.~W.,  2005, \mn@doi [\aap] {10.1051/0004-6361:20041957}, \href
  {http://adsabs.harvard.edu/abs/2005A%26A...434..665R} {434, 665}

\bibitem[\protect\citeauthoryear{{Roxburgh} \& {Vorontsov}}{{Roxburgh} \&
  {Vorontsov}}{2003}]{roxb03}
{Roxburgh} I.~W.,  {Vorontsov} S.~V.,  2003, \mn@doi [\aap]
  {10.1051/0004-6361:20031318}, \href
  {http://adsabs.harvard.edu/abs/2003A%26A...411..215R} {411, 215}

\bibitem[\protect\citeauthoryear{Rumelhart, Hinton  \& Williams}{Rumelhart
  et~al.}{1986}]{rume86}
Rumelhart D.~E.,  Hinton G.~E.,   Williams R.~J.,  1986, \mn@doi [Nature]
  {10.1038/323533a0}, 323, 533

\bibitem[\protect\citeauthoryear{Schmidhuber}{Schmidhuber}{2015}]{schm15}
Schmidhuber J.,  2015, \mn@doi [Neural Networks]
  {10.1016/j.neunet.2014.09.003}, 61, 85

\bibitem[\protect\citeauthoryear{{Scuflaire}, {Th{\'e}ado}, {Montalb{\'a}n},
  {Miglio}, {Bourge}, {Godart}, {Thoul}  \& {Noels}}{{Scuflaire}
  et~al.}{2008}]{scuf08}
{Scuflaire} R.,  {Th{\'e}ado} S.,  {Montalb{\'a}n} J.,  {Miglio} A.,  {Bourge}
  P.-O.,  {Godart} M.,  {Thoul} A.,   {Noels} A.,  2008, \mn@doi [\apss]
  {10.1007/s10509-007-9650-1}, \href
  {http://adsabs.harvard.edu/abs/2008Ap%26SS.316...83S} {316, 83}

\bibitem[\protect\citeauthoryear{{Seaton}}{{Seaton}}{2005}]{seat05}
{Seaton} M.~J.,  2005, \mn@doi [\mnras] {10.1111/j.1365-2966.2005.00019.x},
  \href {http://adsabs.harvard.edu/abs/2005MNRAS.362L...1S} {362, L1}

\bibitem[\protect\citeauthoryear{{Silva Aguirre} et~al.,}{{Silva Aguirre}
  et~al.}{2012}]{silv12}
{Silva Aguirre} V.,  et~al., 2012, \mn@doi [\apj] {10.1088/0004-637X/757/1/99},
  \href {http://adsabs.harvard.edu/abs/2012ApJ...757...99S} {757, 99}

\bibitem[\protect\citeauthoryear{{Silva Aguirre} et~al.,}{{Silva Aguirre}
  et~al.}{2015}]{silv15}
{Silva Aguirre} V.,  et~al., 2015, \mn@doi [\mnras] {10.1093/mnras/stv1388},
  \href {http://adsabs.harvard.edu/abs/2015MNRAS.452.2127S} {452, 2127}

\bibitem[\protect\citeauthoryear{{Thoul}, {Bahcall}  \& {Loeb}}{{Thoul}
  et~al.}{1994}]{thou94}
{Thoul} A.~A.,  {Bahcall} J.~N.,   {Loeb} A.,  1994, \mn@doi [\apj]
  {10.1086/173695}, \href {http://adsabs.harvard.edu/abs/1994ApJ...421..828T}
  {421, 828}

\bibitem[\protect\citeauthoryear{{Ulrich}}{{Ulrich}}{1988}]{ulri88}
{Ulrich} R.~K.,  1988, in {Christensen-Dalsgaard} J.,  {Frandsen} S.,  eds,
  IAU Symposium Vol. 123, Advances in Helio- and Asteroseismology. p.~299

\bibitem[\protect\citeauthoryear{{Valle}, {Dell'Omodarme}, {Prada Moroni}  \&
  {Degl'Innocenti}}{{Valle} et~al.}{2014}]{vall14}
{Valle} G.,  {Dell'Omodarme} M.,  {Prada Moroni} P.~G.,   {Degl'Innocenti} S.,
  2014, \mn@doi [\aap] {10.1051/0004-6361/201322210}, \href
  {http://adsabs.harvard.edu/abs/2014A%26A...561A.125V} {561, A125}

\bibitem[\protect\citeauthoryear{{Verma} et~al.,}{{Verma}
  et~al.}{2014}]{verm14a}
{Verma} K.,  et~al., 2014, \mn@doi [\apj] {10.1088/0004-637X/790/2/138}, \href
  {http://adsabs.harvard.edu/abs/2014ApJ...790..138V} {790, 138}

\makeatother
\end{thebibliography}

\end{document}